\documentclass[aip,reprint]{revtex4-1}

\usepackage{amsmath} 
\usepackage{graphicx}  
\usepackage{verbatim} 
\usepackage{color}   
\usepackage{subfigure} 
\usepackage{hyperref}

\begin{document}

\title{Assessing time series irreversibility through micro-scale trends}

\author{Massimiliano Zanin}
\affiliation{Instituto de F\'isica Interdisciplinar y Sistemas Complejos IFISC (CSIC-UIB),
Campus UIB, 07122 Palma de Mallorca, Spain}

\date{\today}

\begin{abstract}
Time irreversibility, defined as the lack of invariance of the statistical properties of a system or time series under the operation of time reversal, has received an increasing attention during the last decades, thanks to the information it provides about the mechanisms underlying the observed dynamics. Following the need of analysing real-world time series, many irreversibility metrics and tests have been proposed, each one associated with different requirements in terms of e.g. minimum time series length or computational cost. We here build upon previously proposed tests based on the concept of permutation patterns, but deviating from them through the inclusion of information about the amplitude of the signal and how this evolves over time. We show, by means of synthetic time series, that the results yielded by this method are complementary to the ones obtained by using permutation patterns alone, thus suggesting that ``one irreversibility metric does not fit all''. We further apply the proposed metric to the analysis of two real-world data sets.
\end{abstract}

\maketitle

\begin{quotation}
Given a system, or more generally a time series representing the observable dynamics of a system, the first step is usually to try to characterise it through one or more metrics. Among these, tests assessing the irreversible nature of a time series, i.e. whether a time series can or cannot be recognised from its time-reversed version, are gaining attention. Irreversibility stems both from non-linearities and memory in the dynamics, and represents the entropy production of a system out of equilibrium; in short, it can be used to infer information about the physical processes generating a time series, even when these are not directly accessible. We here leverage on a previously proposed metric to estimate the irreversibility of a time series through the concept of permutation patterns, introducing information about the amplitude of the signal and how this changes over time. Most notably, we show that the proposed metric and the original one are complementary, i.e. their relative performance depends on the characteristics of the system under study.
\end{quotation}

\section{Introduction}

The time reversibility of a time series, or more generally of a process, refers to the fact that its statistical properties are invariant under the operation of time reversal; in turns, a time series is said to be irreversible when the result of applying a general function over it changes according to the direction of the arrow of time. Time irreversibility is a fundamental property of non-equilibrium systems, and stems from two properties observed in many real-world systems: the presence of non-conservative forces, i.e. of memory \cite{zwanzig1961memory, puglisi2009irreversible}, and of non-linear dynamics \cite{lawrance1991directionality}. Most importantly, the irreversibility of a time series provides information about the physical mechanism generating it, even when their details are unknown \cite{roldan2010estimating}.

While the concept of time irreversibility is an old one, going back to the philosophy of Aristotle \cite{akih2017beyond}, only recently it has been applied to the study of real-world systems, with an increasing attention being devoted to biological ones. Examples include Parkinson's disease and time series representing the tremors it generates \cite{timmer1993characteristics}; brain dynamics, as e.g. electroencephalographic (EEG) recordings of epileptic patients \cite{van1996time, schindler2016ictal, yao2020permutation} and in other pathologies \cite{zanin2020time}; human gait \cite{orellana2018multiscale, martin2019permutation}; and cardiac dynamics in different conditions \cite{costa2005broken, costa2008multiscale, casali2008multiple, yao2019equal}. Besides biology, time irreversibility has also been studied in, e.g., ecological and epidemiological time series \cite{grenfell1994measles, stone1996detecting}, and finance \cite{ramsey1996time, zumbach2009time, yamashita2017detection}.

Many irreversibility metrics have been developed to support these analyses. To illustrate, the most notable ones include the analysis of consecutive values \cite{ramsey1996time}, the use of symbolic methods \cite{daw2000symbolic}, data compression dictionaries \cite{kennel2004testing}, visibility graphs \cite{lacasa2012time, donges2013testing}, and permutation patterns \cite{zanin2018assessing, martinez2018detection}.

The existence of multiple metrics to detect irreversibility does not only respond to the development of better and more efficient ways of estimating it, but also to the lack of concretion in its definition. As previously stated, irreversibility is defined as any (statistically significant) change in the result of applying a general function on a time series when the arrow of time is reversed; yet, no restriction is imposed on that function. In other words, time irreversibility can appear in any statistical property of the time series, and different metrics have naturally focused on different properties. This may lead to situations in which one time series may be assessed as irreversible by one approach, and as reversible by another one.

In order to understand why an irreversibility metric may fail at correctly classifying a time series, let us introduce the concept of permutation patterns, initially proposed to assess the degree of complexity (or determinism {\it vs.} stochasticity) of time series \cite{bandt2002permutation, zanin2012permutation}. Given a (usually short) window, the corresponding permutation pattern is defined as the order that has to be applied to its elements to sort them - such that, for instance, the pattern associated to values $(8, 4, 5)$ would be $(2, 3, 1)$, as the second element is the smallest one, followed by the third and the first. It has recently been proposed that such patterns can be used to assess irreversibility. In short, the probability of finding a pattern in the original time series should be similar (in a statistic sense) to the probability of finding the same pattern in the time reversed version; if this does not hold true, then that pattern can be used to fix the direction of the time arrow \cite{zanin2018assessing, martinez2018detection}.

\begin{figure*}[!tb]
\begin{center}
\includegraphics[width=0.95\textwidth]{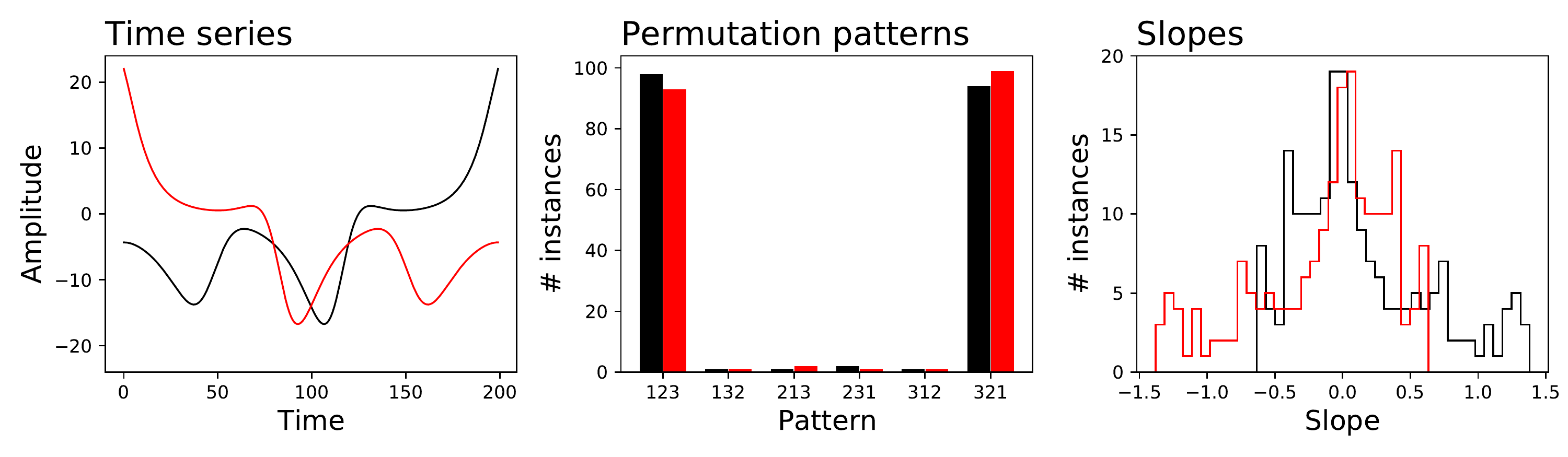}
\caption{Irreversibility in time series. (Left) Example of an irreversible time series (black line) and of its time reversed version (red line). (Centre) Probability distributions of the six permutation patterns extracted from the original time series (black bars) and from the time-reversed version (red bars); note that the two distributions are not different in a statistical significant way. (Right) Distribution of the slopes for the original (black line) and time reversed (red line) time series - see Section \ref{sec:method} for definitions. The two distributions are different in a statistical significant way ($p$-value of $0.0019$, Kolmogorov-Smirnov test).}\label{fig:example}
\end{center}
\end{figure*}

In spite of some important advantages, like their simplicity and reduced computational cost, permutation patterns also present the drawback of disregarding the amplitude of the signal - to illustrate, the two time series $(8, 4, 5)$ and $(16, 8, 10)$ will always result in the same pattern. This well-known fact has important consequences in the estimation of the associated entropy \cite{fadlallah2013weighted, rostaghi2016dispersion}; but further leads to situations in which an irreversible time series may be classified as not irreversible, only because such irreversibility manifests itself in amplitude. As an example, consider the time series represented in Fig. \ref{fig:example} (left panel, black line) and its time reversed version (red line). It can be appreciated by naked eye that it is irreversible, as in fact it is not stationary and displays a clear upward trend. Nevertheless, when the probabilities of the permutation patterns (here considering window sizes of $3$, hence $6$ distinct patterns can appear) are calculated and compared to those of the time reversed version, the difference is not statistically significant - see Fig. \ref{fig:example}, central panel. In other words, triplets of ascending and descending values (respectively corresponding to patterns $(1, 2, 3)$ and $(3, 2, 1)$) appear with approximately the same probability. In fact, irreversibility here does not stem from different pattern probabilities, but from their steepness - i.e. from the amplitude variation within them.

In this contribution we propose an alternative irreversibility metric, based on calculating micro-scale trends, i.e. the slope of a polynomial fit performed over small overlapping windows of the original time series; and on comparing their probability distribution with the one observed in the time-reversed time series. While conceptually similar to the permutation patterns approach, we will show that the inclusion of information about amplitude yields significant advantages when analysing synthetic and real time series, while retaining benefits like being almost parameter-free and of low computational complexity.

\section{Assessing irreversibility in time series}

\subsection{The permutation patterns method}
\label{sec:permp}

For the sake of completeness, we here synthesise how irreversibility can be detected by means of permutation patterns. Note that several methods have independently been proposed in the last years \cite{zanin2018assessing, martinez2018detection, li2019time, yao2019quantifying}; while the underlying concept is the same, namely extracting permutation patterns and comparing their frequency, details about how the statistical significance of results is assessed vary. While we here use the method proposed in \cite{zanin2018assessing}, the reader should be aware of the available alternatives \cite{martinez2018detection, li2019time, yao2019quantifying}.

Given a time series $X = \{ x_1, x_2, \ldots, x_N \}$ composed of $N$ values, this is divided into $N - \delta + 1$ overlapping windows of length $\delta$, such that the $i$-th window is defined as $w_i = \{ x_i, \ldots, x_{i + \delta - 1} \}$. Values composing each window are then sorted from smaller to larger, and the permutation $\pi_i$ needed to perform this sorting is extracted. To illustrate, consider a time series $X = \{4, 7, 5, 6, \ldots\}$; for $\delta = 3$, $w_1$ is composed of values $\{4, 7, 5\}$, and the corresponding permutation pattern is $\pi_1 = (1, 3, 2)$. $\pi_i$ thus represents how values should be reordered to sort them, and hence the structure by them created \cite{bandt2002permutation, zanin2012permutation}. 

Irreversibility can be assessed by noting that the distribution of the permutation patterns probability $p(\pi)$ should be the same in the original and time reversed time series. Such similarity can be assessed through the Jensen-Shannon divergence, a symmetric version of the Kullback-Leibler divergence that measures the similarity between two probability distributions \cite{grosse2002analysis}. Finally, a test based on surrogate time series can be performed to check the statistical significance of the difference between the two distributions, and eventually obtain a $p$-value.

\subsection{Introducing amplitude: the micro-scale trends method}
\label{sec:method}

As in the previous case, let us suppose a time series $X = \{ x_1, x_2, \ldots, x_N \}$ composed of $N$ values, also divided into $N - \delta + 1$ overlapping windows of length $\delta$, such that the $i$-th window is defined as $w_i = \{ x_i, \ldots, x_{i + \delta - 1} \}$. Afterwards, a least squares polynomial fit of degree $d < \delta$ is applied to each window, and the highest power coefficient $a$ is extracted. Finally, two probability distributions of $a$, for the original and time-reversed time series, are extracted and compared through a Kolmogorov-Smirnov test.

The behaviour of this test can be clarified considering the simplest situation of $\delta = 2$ and $d = 1$. In this case, any window $i$ will be composed of two values $\{ x_i, x_{i+1} \}$, the fit will be a linear one, and the coefficient $a_i$ will correspond to the slope, i.e. $a_i = x_{i+1} - x_i$. The two probability distributions $P_a$ and $P_{a^t}$ then represent the distribution of the discrete derivative of the original and time reversed time series, or, in other words, of their slopes. If these are different in a statistically significant way, as evaluated by the Kolmogorov-Smirnov (K-S) test, then the irreversibility hypothesis is accepted.
This process is illustrated in Fig. \ref{fig:example} right panel, which depicts the two distributions $P_a$ (black line) and $P_{a^t}$ (red line) for the time series represented in the left panel. The instances of sharp increases in the original time series (slope $> 0.5$) are not seen in the time reversed version, as these become sharp decreases (slope $< -0.5$). This asymmetry in the distribution of slopes then highlights the presence of an irreversible process.

The similarities and differences with the permutation patterns approach are easy to visualise. In the case of $\delta = 2$ and $d = 1$, this test is equivalent to performing an analysis based on permutation patterns of length $2$; yet, we here consider the slope, or excursion between consecutive values, instead of its sign only. Similarly, the case of $\delta = 3$ and $d = 1$ is a linear fit between three consecutive values; the middle one is then disregarded, and this is equivalent to considering the permutation pattern created by values $x_i$ and $x_{i+2}$.

For the sake of simplicity, in this contribution we only consider the cases for $\delta = 2 \ldots 10$ and $d = 1$, i.e. linear fits on windows of small length. Nevertheless, the method here proposed can easily be extended to more complex situations, which can yield richer views of the dynamics. First of all, polynomials of any order can be extracted; the distributions $P_a$ and $P_{a^t}$ will then represent the dominant trend in the windows. Secondly, as customary in the evaluation of permutation patterns, one can add lags in the reconstruction of the windows, such that $w_i = \{ x_i, x_{i+\gamma}, x_{i+2\gamma}, \ldots \}$. Finally, any transformation of the original time series can be used; for instance, one could create a new time series with the standard deviation of each window, thus representing the evolution of the dispersion of values, for then evaluating the irreversibility of this new time series.

\section{Evaluation on synthetic time series}
\label{sec:synth}

In order to evaluate the capacity of this method to detect irreversible dynamics, we firstly test it with a set of time series created by the following standard processes:

\begin{itemize}

\item A Gaussian noise of zero mean and unitary standard deviation.

\item The Arnold Cat map, defined as: $x_{n+1} = ( x_n + y_n) mod (1)$, $y_{n+1} = ( x_n + 2y_n ) mod (1)$. The analysed time series corresponds to the evolution of the $x$ variable.

\item The Ornstein-Uhlenbeck process, i.e. a mean-reverting linear Gaussian process $T$ \cite{weiss1975time}.

\item The Generalised Autoregressive Conditional Heteroskedasticity (GARCH) model \cite{bollerslev1986generalized}, defined as: $x_t = \sigma_t z_t$, with $\sigma ^2 _t = \alpha ^* ( 1 + \sum _{i=1} ^3 2^{-i}) x ^2 _{t-i} + \sum _{i=1} ^3 2^{-j} \sigma^2 _{t-j}$, and $z_t$ being independent random numbers drawn from an uniform distribution $\mathcal{U}(0, 1)$. $\alpha ^*$ (here set to $1.5$) is a parameter controlling the strength of the time dependence between present and past values of $x$, and hence the irreversibility of the time series.

\item The Henon map (defined as $x_{n+1} = 1 + y_n - a x_n^2$, $y_{n+1} = b x_n$, with $a = 1.4$ and $b = 0.3$) and the logistic map ($x_{n+1} = a x_n (1 - x_n)$, with $a = 4.0$). Both maps are dissipative systems, and are by definition irreversible \cite{mori2013dissipative}.

\item The Lorenz chaotic system, a continuos system defined as $x' = \sigma(y - x)$, $y' = x(\rho - z) - y$, and $z' = xy - \beta z$, with $\rho = 28$, $\sigma = 10$ and $\beta = 8/3$. The system has been solved with integration steps of $dt = 0.01$. The three variables $x$, $y$ and $z$ are here analysed independently. 

\end{itemize}

Both the Gaussian noise, the Arnold Cat map (an example of a conservative chaotic map) and the Ornstein-Uhlenbeck process generate reversible time series, while all others are irreversible.

\begin{figure*}[!tb]
\begin{center}
\includegraphics[width=0.95\textwidth]{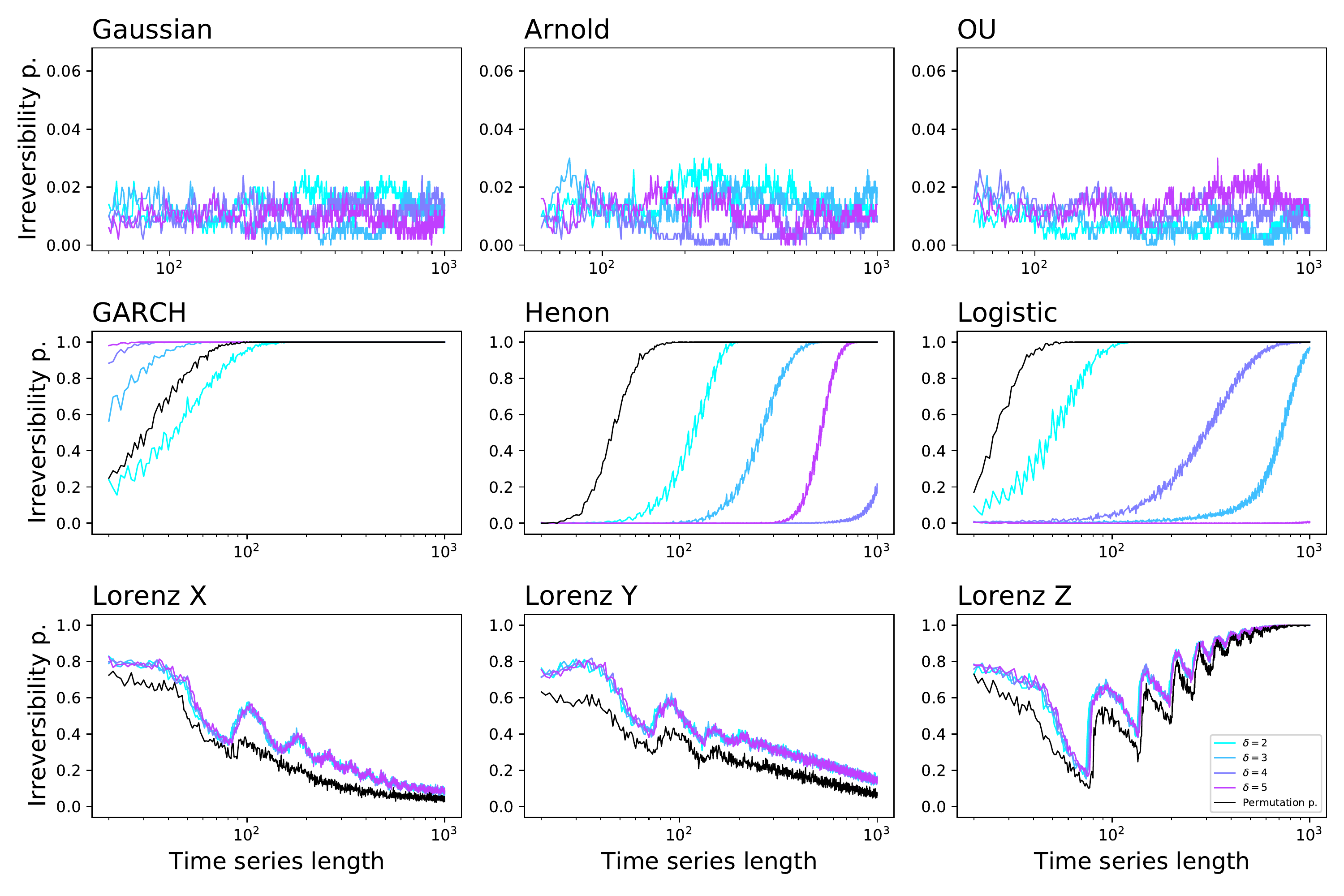}
\caption{Evolution of the probability of finding a statistically significant time series as a function of its length, for the processes described in Sec. \ref{sec:synth}. Lines with color between cyan and purple correspond to the proposed method, for $\delta$ between $2$ and $5$; black lines correspond to the permutation patterns method.}\label{fig:results}
\end{center}
\end{figure*}

Fig. \ref{fig:results} reports the evolution of the probability of finding an irreversible time series as a function of its length, for the nine systems considered, and for $\delta = 2 \ldots 5$ and $d = 1$. As a reference, the black lines of the same figure report the results obtained with the permutation patterns approach described in Sec. \ref{sec:permp}. It can be appreciated that the proposed metric is able to correctly detect the presence or absence of irreversibility, provided enough data (i.e. long enough time series) are available. Also, results are generally similar to the ones yielded by the permutation patterns approach. Still, several interesting differences can also be observed. Specifically, the proposed method requires smaller time series in the case of the GARCH model, and is slightly better in detecting the irreversibility of the Lorenz system. On the other hand, it requires longer time series (i.e. it is less sensitive) in the cases of the Henon and logistic maps.

\begin{figure*}[!tb]
\begin{center}
\includegraphics[width=0.95\textwidth]{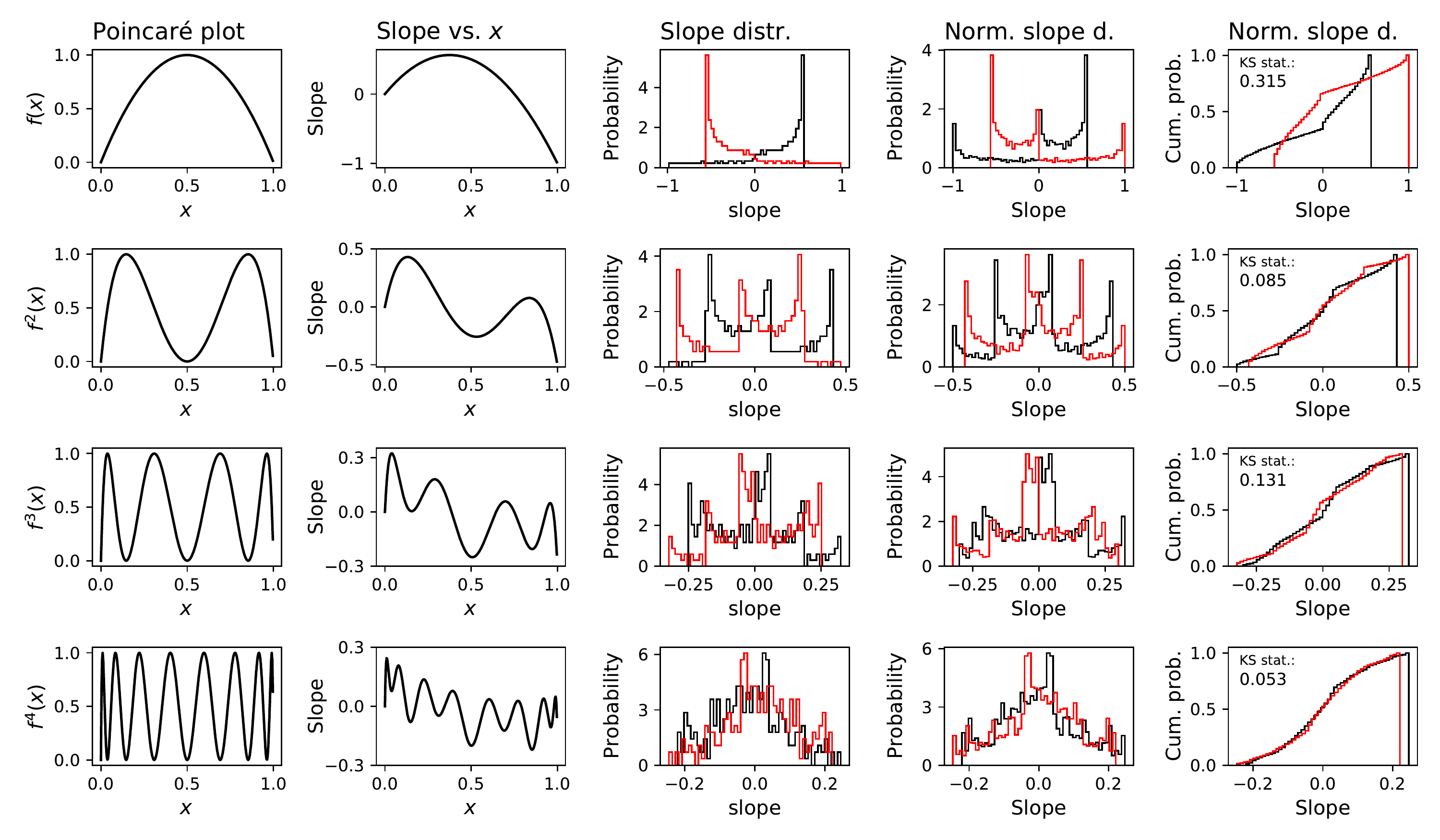}
\caption{Analysis of the irreversibility of the logistic map. From top to bottom, rows correspond to $\delta = 2$, $3$, $4$ and $5$. From left to right, columns correspond to: i) Poincar\'e plot of the map, depicting the evolution of $f^{\delta - 1}(x)$ as a function of $x$. ii) Slope of the linear fit of values $(x_t, \ldots, x_{t+\delta})$ as a function of $x$. iii) Histogram of the probability distribution of the slopes. iv) Histogram of the probability distribution of the slopes, when values are normalised according to the stationary distribution of $x$. v) Cumulative probability distributions of the normalised slopes; the text inside each panel reports the statistics of the K-S test assessing the equality of both distributions. In the third, fourth and fifth columns, black and red lines respectively correspond to the original and time reversed time series.}\label{fig:Logistic}
\end{center}
\end{figure*}

Fig. \ref{fig:results} also indicates that the role of $\delta$, i.e. of the length of the windows on which the fit is calculated, is a complex one. On one hand, in the case of the Lorenz chaotic system, different $\delta$s yield the same result. On the other hand, results differ substantially for the GARCH, Henon and logistic time series; the test is more sensitive for large $\delta$s in the former case, while the opposite can be observed for the two chaotic maps. Such heterogeneity can easily be explained by considering the nature of these time series. Data generated by GARCH are not stationary; larger $\delta$s thus allow to filter out the local noise and extract the main trend. Conversely, time series generated by the Henon and logistic maps are bounded and stationary; calculating the slope of a linear fit over long windows is effectively smoothing out the dynamics, and the slope actually becomes zero in the limit of infinitely long windows, thus erasing differences between the original and time reversed time series. Exceptions are nevertheless present: for instance, the test is more sensitive in the case of the logistic map for $\delta = 4$ than for $\delta = 3$. In order to understand this behaviour, Fig. \ref{fig:Logistic} reports several graphs associated to the map, for $\delta$ varying between $2$ (top row) and $5$ (bottom row). When one considers the probability distributions of the slopes for $\delta = 3$ and $4$, normalised according to the stationary distributions of $X$, both present differences between the original (black lines) and time reserved (red lines) data; these differences are nevertheless centred around zero in the case of $\delta = 4$, thus yielding a larger K-S statistics and a smaller $p$-value. In synthesis, these results highlight that, whenever the characteristics of the underlying dynamics are not known a priori, one should compare the results corresponding to different $\delta$s to achieve an optimal detection.

\begin{figure*}[!tb]
\begin{center}
\includegraphics[width=0.95\textwidth]{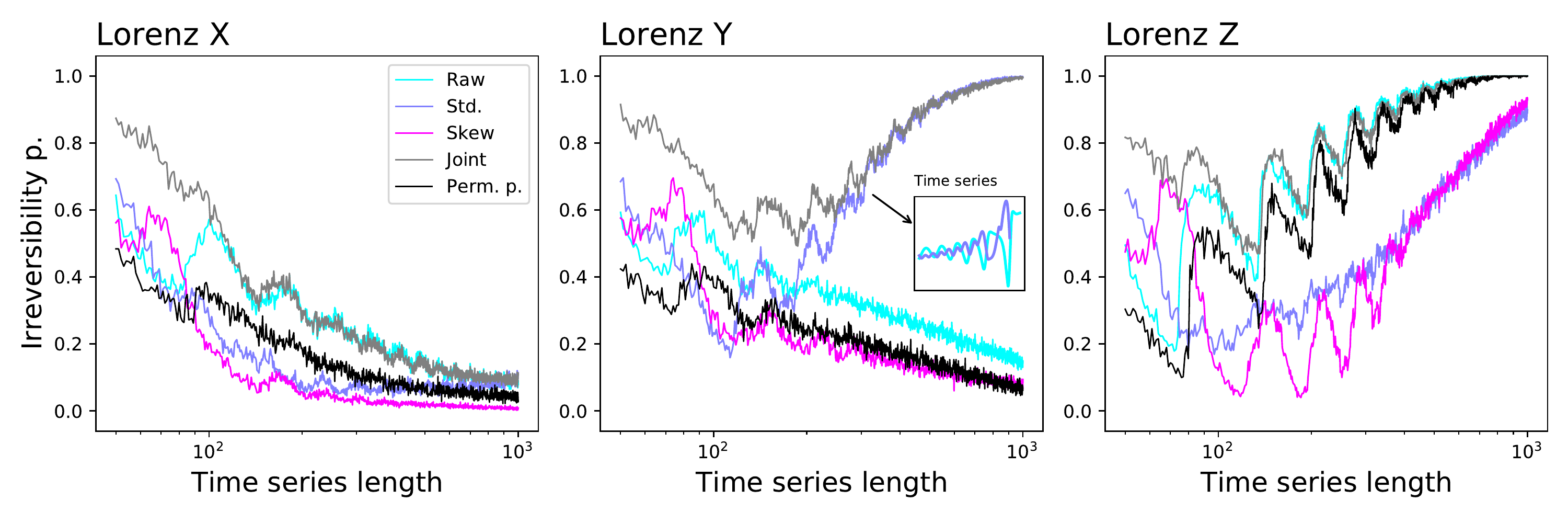}
\caption{Evolution of the probability of finding a statistically significant time series as a function of its length, for the three channels of a Lorenz system, and considering: the raw time series (cyan lines), the time series of standard deviation (blue) and skewness (purple), the joint test over the previous three time series (grey), and by using the permutation pattern test (black). The inset in the central panel shows an example of such time series, and specifically a raw one (cyan) and the corresponding standard deviation time series (blue).}\label{fig:Lorenz}
\end{center}
\end{figure*}

As shortly introduced in Sec. \ref{sec:method}, the proposed method can be applied to any modification of the original time series. To illustrate, a new time series $y_t$ can be created by calculating the second (standard deviation) and third (skewness) central moments of the sub-time series $\{x_t, \ldots, x_{t+\Delta}\}$ (note that $\Delta \neq \delta$, and that usually $\Delta \gg \delta$); the method can then be applied to the new time series $y$. This allows detecting situations in which, for instance, the time series is alternating between upward and downward movements with an increasing (or decreasing) frequency; while the same number of positive and negative slopes will appear, and hence no irreversibility will be detected in the raw time series, the change in frequency will reduce (or increase) the deviation from the mean, yielding an irreversibility in the time series of the standard deviation. This idea is applied in Fig. \ref{fig:Lorenz}, depicting the evolution of the irreversibility in the Lorenz system for the raw, standard deviation and skewness time series (for $\Delta = 20$). It can be appreciated that, while in most cases the derived time series underperform, calculating the irreversibility over the time series of the standard deviation for the Y channel completely reverses the result. The reason is readily identifiable by looking at an example of the corresponding time series (see inset in the central panel): the raw time series (cyan line) increases in amplitude while maintaining a constant average, which translates to a clear non-stationarity (and hence, irreversibility) of the time series of the standard deviation (blue line). When a priori information about the time series is not available, a solution may entail performing the test on the three time series, for then accepting the original time series as irreversible if any of the three tests yielded a statistically significant result after correcting for multiple comparisons. This is illustrated in Fig. \ref{fig:Lorenz} by the grey lines, depicting the evolution of the fraction of irreversible time series when the three tests are combined ($\alpha = 0.01$ with a Bonferroni correction).

\section{Evaluation on real-world time series}

We then move to the evaluation of the proposed method in real-world time series, and specifically consider two examples: the analysis of brain electroencephalographic (EEG) data, and time series representing the evolution of delays in the air transport system.

\subsection{Brain electroencephalographic data}

Given that the irreversibility of a dynamical system is related to its entropy production and to its performance as a thermal machine, it is not surprising to find numerous applications of this concept to the study of the human brain \cite{van1996time, schindler2016ictal, yao2020permutation, zanin2020time}. Specifically, if a disease or condition is impairing the self-organising capabilities of the brain, this should reflect in an abnormal (either higher or smaller) time irreversibility, and the latter could therefore be used as a marker of the former. Previous works have nevertheless shown that irreversibility is not an easy to detect property of brain signals, and that long time series are usually required \citep{martinez2018detection, zanin2020time}. This may preclude the use of this property in the study of brain dynamics developing on short time scales, as for instance the response to stimuli.

As a first real-world test, we here apply the proposed irreversibility metric to a data set of electroencephalographic (EEG) recordings, comprising both control subjects and patients suffering from alcoholism \citep{zhang1995event, cao2014disturbed, zanin2021uncertainty} and available at \url{https://archive.ics.uci.edu/ml/datasets/EEG+Database}. Each recording corresponds to the execution of a standard object recognition task \citep{snodgrass1980standardized}, and includes $64$ time series (i.e. one for each of the $64$ electrodes) of $256$ elements (i.e. one second of brain activity). Note the reduced length of these time series, which would preclude obtaining statistically significant results with the method proposed in \citep{zanin2018assessing} - see also \citep{zanin2020time} for a more complete analysis. On a positive side, a large number of trials are available, specifically $4,024$ for control people and $7,033$ for patients.

\begin{figure*}[!tb]
\begin{center}
\includegraphics[width=0.99\textwidth]{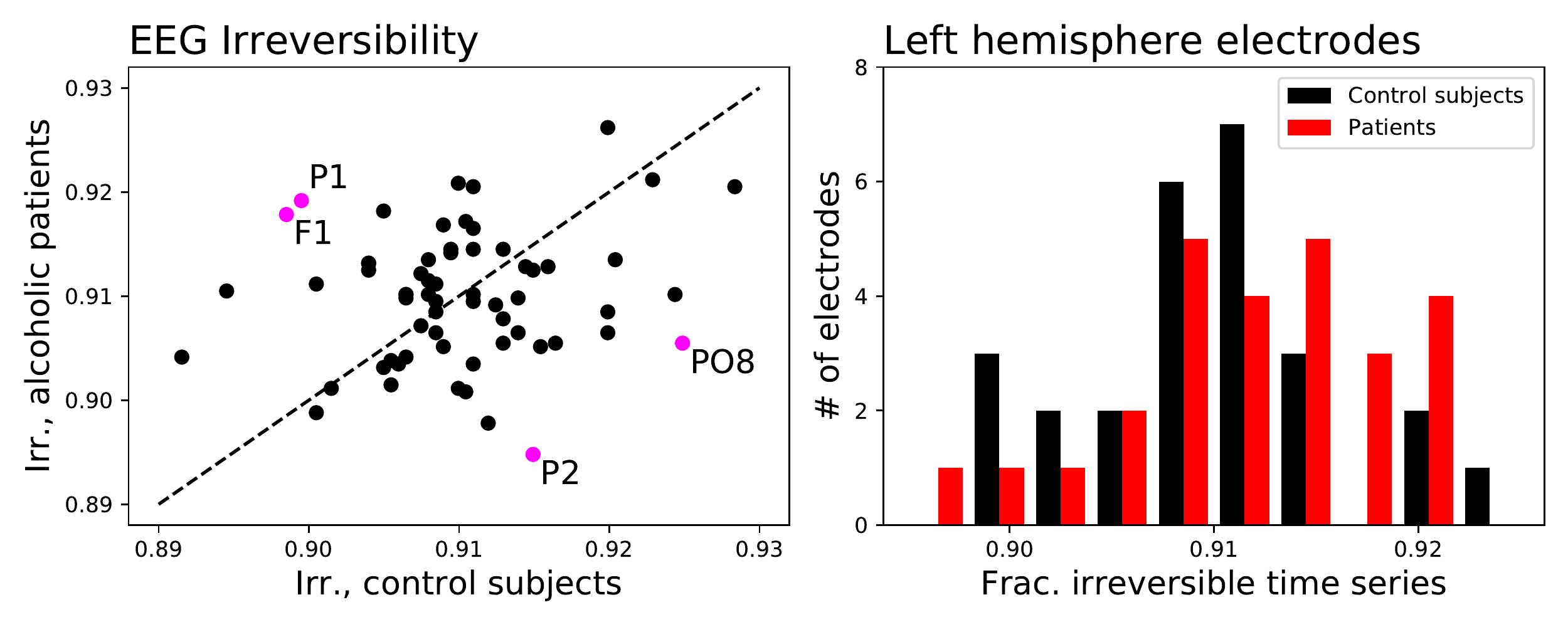}
\caption{Irreversibility of electroencephalographic (EEG) recordings. (Left) Scatter plot of the fraction of patients' recordings detected as irreversible, as a function of the same fraction for control subjects. Each point represents an electrode; magenta points further indicate the electrodes for which the difference between the two fractions is statistically significant, according to a binomial test and for $\alpha = 0.01$. (Right) Histogram of the number of electrodes in the left hemisphere as a function of the corresponding fraction of irreversible time series, for control subjects (black columns) and patients (red columns). }\label{fig:EEG}
\end{center}
\end{figure*}

Fig. \ref{fig:EEG} Left reports a scatter plot of the fraction of time series that were detected as irreversible by the proposed method for patients, as a function of the fraction of irreversible time series in control subjects, with each point representing a different EEG channel. In other words, channels laying close to the main diagonal, represented by the black dashed line, display a similar degree of irreversibility both in patients and control subjects. It is firstly worth noting that the fraction of irreversible time series is globally quite high, i.e. between $89\%$ and $93\%$; obtaining such strong signals using a permutation patterns-based method would require time series $100$ times longer \citep{zanin2020time}. Secondly, interesting differences between the two groups can be observed, and specifically that time series of P1 and F1 electrodes are more irreversible in patients, and of P2 and PO8 in control subjects; alcoholic patients seem thus to have a more irreversible dynamics in the left hemisphere and a more reversible one in the right hemisphere. This is further confirmed by Fig. \ref{fig:EEG} Right, depicting an histogram of the number of electrodes in the left hemisphere as a function of the fraction of irreversible time series, both for control subjects (black columns) and patients (red columns); here again time series corresponding to electrodes in the left hemisphere of patients are associated with higher irreversibility than patients' ones. While the relationship between cerebral laterality and alcoholism has long been studied \citep{london1987cerebral, ellis1989alcoholism, akshoomoff1989block, mcnamara1994markers, zhu2014analysis}, to the best of our knowledge this is the first time irreversibility of brain dynamics is introduced in the picture.

\subsection{Air transport delay data}

As a second real-world example, we here consider time series representing the evolution of delays at the $30$ largest European airports. Relatively few works have studied the dynamics of delays from the point of view of statistical physics, and this in spite of their importance for the cost-efficiency \cite{cook2011european}, safety \cite{duytschaever1993development}, and environmental impact \cite{carlier2007environmental} of this transportation mode. One may prima facie expect delays to be random and independent events - as e.g. when one aircraft experiences some technical problems, and has to delay the take off; as such, the time series representing the evolution of the average delay should resemble a random process, and hence not be irreversible. On the other hand, if delays are not independent (as e.g. when the delay of one flight is caused by the late arrival of the previous one), the memory present in the system will result in an irreversible dynamics. Note that discriminating between these two cases is not just an academic exercise, as delays of the first type are unpredictable, while on the contrary those of the second type are the result of inefficiencies in the system and can be avoided.

Air traffic data have been extracted from the EUROCONTROL's R\&D data archive \cite{RandDArchive}, a large and freely available repository of information about the European airspace and all commercial flights crossing it. The data set covers four years, from 2015 to 2018, with four months being available for each year (March, June, September and December). All flights landing at the $30$ largest European airports (according to their number of passengers in 2015) have been extracted, for then calculating their delay at landing as the difference between the actual and the scheduled landing times. All flights that have landed at a given airport and in a given hour have then been aggregated, to obtain a time series of average hourly delay at each airport.

\begin{figure*}[!tb]
\begin{center}
\includegraphics[width=0.95\textwidth]{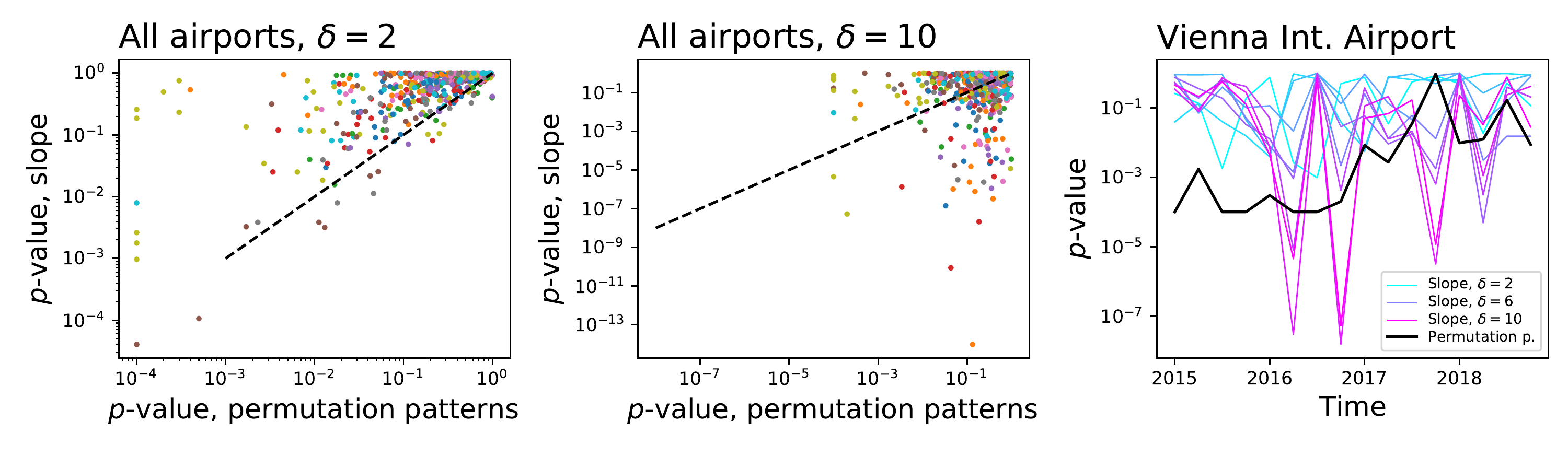}
\caption{Irreversibility of airport delay time series. Left and central panels report the $p$-value of the irreversibility of the time series of each airport (see main text for definitions), as a function of the $p$-value obtained by the permutation patterns method \cite{zanin2018assessing}, for $\delta = 2$ (left panel) and $\delta = 10$ (central panel). (Right) Monthly evolution of the $p$-value of the irreversibility for Vienna International Airport, according to the permutation patterns method (black line) and the proposed method (from $\delta = 2$, cyan lines, to $\delta = 10$, magenta lines). }\label{fig:Delays}
\end{center}
\end{figure*}

Each of the $16$ available months have been analysed independently, and the irreversibility has thus been calculated over time series of length $720 - 744$ (i.e. $24$ values per day, $30$ or $31$ days depending on the month). When including $30$ airports, a total of $480$ time series have been analysed. Fig. \ref{fig:Delays} reports the scatter plot of the $p$-values obtained with the proposed method, as a function of the $p$-value yielded by the permutation patterns approach, for $\delta = 2$ (left panel) and $\delta = 10$ (central panel). It can be observed that, while for $\delta = 2$ the $p$-values obtained by the former method are usually larger than those corresponding to the latter one, the opposite happens for $\delta = 10$. Specifically, it is possible to observe a cluster of points for which the $p$-value yielded by the permutation pattern is not statistically significant ($\approx 10^{-1}$), while the one yielded by the trends method is around $10^{-5} - 10^{-7}$.
The right panel of Fig. \ref{fig:Delays} finally report the evolution of the $p$-value of the irreversibility, as yielded by both methods, for the Vienna International Airport (i.e. the airport with the highest average irreversibility). While the permutation pattern approach yields increasing $p$-values, thus suggesting less systemic delays, the proposed method points towards the presence of non-random delays in June and December 2016, December 2017 and June 2018.
In short, it can be observed that both methods are not equivalent, but actually complementary. The proposed method is able to detect irreversibility for some time series that are identified as reversible through permutation patterns; and in this case, it is more sensitive to irreversibility for large values of $\delta$, as previously seen in the case of the GARCH model.

\section{Discussion and conclusions}

The recently increasing interest in the concept of irreversibility has been followed closely by an increase in the number of metrics designed to detect irreversibility in real-world time series. These approaches do not represent differential increments, e.g. minor improvements in the computational cost, but instead yield complementary views on the concept of irreversibility, as the concept itself is only loosely defined. A number of these metrics leverage on the concept of permutation patterns and have been independently proposed by several research groups \cite{zanin2018assessing, martinez2018detection, li2019time, yao2019quantifying}, being increasingly applied to the study of experimental data sets. We nevertheless here show that permutation patterns alone cannot identify irreversibility in some pathological cases, due to the fact that they disregard information about the amplitude (or magnitude) of the values composing the time series. In other words, from the permutation patterns point of view, the sequences $(0, 1, 2)$, $(0, 10, 20)$ and $(100, 101, 102)$ are identical. In this contribution we thus propose an alternative method, based on evaluating the changes in the amplitude of the time series' values, through micro-scale regressions of the time series and of transformations of it. 

The evaluation of the proposed method through synthetic time series depicts an interesting picture: its performance strongly depends on the dynamical system under analysis. To illustrate, the proposed method is able to detect the irreversible nature of the Y channel of a Lorenz system, something not achieved by a permutation pattern approach (see Fig. \ref{fig:Lorenz}); yet, the latter is substantially more efficient at detecting irreversibility in Henon and logistic maps (see Fig. \ref{fig:results}). In other words, and consistently with its definition, the proposed method ought to be used when the amplitude of the signal is not constant, for instance due to local non-stationarities. This yields major benefits in the case of the analysis of real-world time series, which are not necessarily stationary. To illustrate, the proposed method was able to detect the irreversibility of brain EEG recordings even for time series composed of only 256 points, while previous attempts required substantially longer recordings \cite{zanin2020time}.

On the other hand, it is also important to highlight some limitations of the present approach. First of all, while it yields better results in the case of some dynamical systems, it is not clear when this is the case; in other words, we cannot provide a decision algorithm that suggests the best test to be used given one time series - beyond, of course, the brute-force approach of trying all possible algorithms. This is not only limited to the proposed approach, but is instead an open research question. Secondly, the proposed approach includes some parameters that have to be tuned to maximise the sensitivity of the irreversibility test, including the sub-window length $\delta$ and the use of transformed time series. Note that their tuning is more complex than e.g. tuning the embedding dimension of permutation patterns, as, provided enough data are available, higher embedding dimensions are usually better. On the contrary, and as shown in Figs. \ref{fig:results} and \ref{fig:Logistic}, increasing $\delta$ may lead to a reduced statistical significance.

\begin{acknowledgements}
This project has received funding from the European Research Council (ERC) under the European Union's Horizon 2020 research and innovation programme (grant agreement No 851255).

M.Z. acknowledges the Spanish State Research Agency, through the Severo Ochoa and Mar\'ia de Maeztu Program for Centers and Units of Excellence in R\&D (MDM-2017-0711).
\end{acknowledgements}

\section*{Data availability}
The data that supports the findings of this study are available within the article.

\bibliography{Slope}

\end{document}